\documentclass[article]{JHEP}
\usepackage{amsmath,epsfig}
\usepackage{latexsym}
\usepackage{amssymb,amsfonts,bm}
\usepackage{feynmf}
\usepackage{url}
\usepackage{amssymb,amsfonts}
\relax
\def\be{\begin{equation}}
\def\ee{\end{equation}}
\def\bea{\begin{eqnarray}}
\def\eea{\end{eqnarray}}
\def\be{\begin{equation}}
\def\ee{\end{equation}}

\def\ie{{\it i.e.}~}
\def\eg{{\it e.g.}~}

\newcommand\fverb{\setbox\pippobox=\hbox\bgroup\verb}
\newcommand\fverbdo{\egroup\medskip\noindent%
                        \fbox{\unhbox\pippobox}\ }
\newcommand\fverbit{\egroup\item[\fbox{\unhbox\pippobox}]}
\newbox\pippobox

\def\e{\epsilon}

\def\m{\mu}

\def\a{\alpha}

\def\sp{\;\;\;,\;\;\;}

\def\ba{\begin{eqnarray}}
\def\ea{\end{eqnarray}}

\def\de{\partial}

\def\hre#1#2{\href{http://arxiv.org/abs/#1/#2}{[ArXiv:#1/#2]}}

\def\sla{\raise.15ex\hbox{$/$}\kern-.57em}

\def\tr{{\rm tr}}

\def\cC{{\cal C}}

\def\cF{{\cal F}}

\def\cH{{\cal H}}

\def\cM{{\cal M}}
\def\cN{{\cal N}}
\def\cO{{\cal O}}

\def\cT{{\cal T}}

\def\ap{{\alpha^\prime}}

\title{Non-perturbative and Flux superpotentials for
Type I strings on the ${\bf Z}_3$ orbifold}

\author{\large Massimo Bianchi$\ ^1$ and  Elias
Kiritsis$\ ^{2,3}$\\
~\\
$^1~$Dipartimento di Fisica, \ Universit{\`a} di Roma \ ``Tor
Vergata''
$~$\\
I.N.F.N.\ -\ Sezione di Roma \ ``Tor Vergata''
$~$\\
Via della Ricerca  Scientifica, 1 - 00133 \ Roma, \ ITALY
\\
$~$\\
$^2~$CPHT, UMR du CNRS 7644, Ecole Polytechnique, CNRS,
91128 Palaiseau, FRANCE\\
$~$\\
$^3~$Department of Physics, University of Crete, 71003 Heraklion,
GREECE}

\preprint{\hepth{0702015}\\
CPHT-RR005.0107\\
ROM2F/07/04}      % OR:
\abstract{Non-perturbative effects are studied for Type I strings
on the ${\bf Z}_3$ orbifold with Chan-Paton symmetry broken by
Wilson lines. Generalizing previous analyses that have focussed on
(bi)fundamentals, it is  argued that (anti)symmetric
representations of the resulting gauge group play a decisive role
in generating an ADS-like superpotential in this and related
cases.  Non-perturbative corrections in the closed string moduli
are only allowed if properly dressed with open string fields
charged under the anomalous $U(1)$ of the orbifold. A general
discussion of instanton effects in SYM and string theories is
given. Non-perturbative superpotentials induced by both $ED5$'s
and $ED1$'s are analyzed. The superpotential generated by closed
string fluxes, {\it viz.} Scherk-Schwarz shifts (torsion), R-R
3-form flux as well as non-geometric fluxes is derived. Some
preliminary comments on the compatibility of the two kinds of
superpotentials and the issue of moduli stabilization are
presented.}

\begin{document}

{}{}{}{}{}{}{}{}{}{}{}{}{}{}{}{}{}{}{}{}{}{}

\section{Introduction and conclusions}

Vacuum configurations with open and  unoriented strings
\cite{opensys} have received attention thanks to their
remarkable phenomenological properties \cite{unorev}.  In such
vacua the standard model gauge group is realized on stacks of
intersecting or magnetized (co-isotropic) $D$-branes that give
rise to massless chiral fermions at their intersections or as a
result of the degeneracy of Landau levels \cite{intermagn}. Closed
strings propagate in the bulk and mediate gravity and several
other unobserved interactions through the exchange of scalar
moduli (together with their superpartners).

The strong phenomenological appeal of orientifold vacua is mostly
due to the fact that ``bottom-up" approaches, \cite{bu}, to realizing the
Standard Model spectrum and interactions can be more or less implemented.
In such approaches, one can assemble a local brane
configuration with the required phenomenological properties, and
postpone, global consistency conditions like tadpole cancellation.
Indeed, in many cases this leads to configurations that can be
eventually upgraded to {\it bona fide} vacua with a high degree of
success\footnote{In the large scale computerized searches of
\cite{schell,adks}, local configurations that satisfy basic BCFT
requirements have $\sim$1\% probability to be completed into full vacua,
by adjusting the hidden sector.}

 Turning on internal fluxes both in the  closed
\cite{closedflux} and open string \cite{openflux} sector one can
achieve (supersymmetric) moduli stabilization in AdS space and
then try to uplift the system to a metastable De Sitter vacuum
with a tiny cosmological constant \cite{KKLT}. The last step is
very delicate and requires taking carefully into account all
perturbative and non-perturbative contributions to the potential
simultaneously. Some of them are under control in full-fledged
string theory (such as magnetic fluxes and NS-NS geometric and
non-geometric fluxes), other (R-R fluxes) can be studied in
the diluted flux approximation or in an effective supergravity
description.

It is widely  appreciated that a satisfactory implementation of
supersymmetry breaking and moduli stabilization in string theory
may not forgo a complete understanding of non-perturbative effects
associated to (Euclidean) branes wrapping internal cycles
\cite{wsinst, BBStrom, stringinst}.

An omnipresent, as well as crucial feature of orientifold vacua,
is the existence of  (potentially anomalous) $U(1)$ factors in the
gauge group. Mixed anomalies of such factors have been shown to
cancel both in 6D \cite{6sagn} and 4D vacua \cite{ib} by variants
of the Green-Schwarz mechanism. However, it was subsequently
appreciated, that there is a richer structure associated to
anomalous U(1)'s that is not tied only to 4D anomalies
\cite{ib1,art,anasta,abdk}. The presence of anomalous U(1), and
their couplings require global knowledge of the orientifold
ground-state, and in general cannot be successfully treated in
local brane configurations\footnote{Some exceptions to this
statement exist, in special oriented ground-states where isolated
stacks can appear, \cite{hv}. However, such examples remain to be
seen whether they survive the cancellation of tadpole conditions.}
 Moreover, since the anomalous
$U(1)$ symmetries are inextricably related to isometries of the
manifold of closed string moduli, they are an integral part of any
attempt to generate potentials and stabilize moduli. In initial
stabilization setups, the role of anomalous U(1)'s was ignored.
Their presence however is important in non-perturbative
superpotentials, as the associated gauge invariance constraints
the type of terms that appear, as it was first appreciated in
\cite{dma} and recently studied in \cite{stringinst}. However, all
related studies so far have been ``local" and have avoided the
global constraints of orientifolds.

The study of string instanton effects  is still in its infancy.
Several situations, especially with extended supersymmetry have
been analyzed in the past (see \cite{e-instrev} for a review)
drawing mostly on non-perturbative dualities of string vacua with
extended supersymmetry. However, the direct study of string
instanton effects still remains an anecdotal subject (modulo some
recent efforts \cite{instfromstr},\cite{stringinst}).

 In this paper we initiate their study for the simple case of the
${\bf Z}_3$ orbifold \cite{chiras}. As we will see,
(anti)symmetric representations of the Chan-Paton group will play
a crucial role in these cases, generalizing previous analyses that have
so far only considered (bi)fundamentals.

Depending on the stack of branes under  consideration there are
essentially two distinct classes of 'instantonic' branes. The
first class involves branes that form bound states at threshold
with the previous stack such as when they share 4 N-D directions.
The second includes branes that share 8 N-D directions and
accommodate only a chiral fermion at the intersection. The former
are akin to gauge instantons \cite{instfromstr}. The prototype is
the $D3$, $D(-1)$ system studied from various perspectives over
the years \cite{instfromstr}. The latter have only very recently
attracted some attention \cite{stringinst} and may eventually
enjoy a field theory description in terms of octionionic
instantons or hyper-instantons \cite{hyperinst}.
We will not
pursue this point any further here but we will identify the role
of this kind of stringy instantons in the vacuum we consider.

Including closed string fluxes we will  have a complete picture of
the full (super)potential and we will attempt a preliminary
analysis of the resulting moduli stabilization problem.

In this paper we improve on previous analyses of instanton effects
and non-perturbative superpotentials in several directions:

\begin{itemize}

\item  We provide a clear classification of all $D$-brane
instanton effects for Type I strings on the ${\bf Z}_3$ orbifold.

\item  We provide the correct form of various vertex  operators
corresponding to instanton modes, in particular the fermionic zero
modes that are relevant.

\item We include the effect of (anti) symmetric representations (eg. 6 of SU(4))
that are carefully avoided in previous works.

\item We identify concrete rigid cycles (the combination of the exceptional divisors), In
toroidal models where concrete efforts have been undertaken, all
cycles are unfortunately sliding ones.

\item We give a precise identification of the consistency
conditions (Bianchi identities)  $dG_1 = F_2$ that constrain
$ED$-string wrapping.

\item We identify the (U(1)-charged) prefactors of instanton contributions to superpotential in two ways: by an instanton
zero-mode counting and by U(1) neutrality (obviously the two are
related)

\item We finally give an expression for the superpotential with geometrical fluxes
($F_3$ and torsion) as well as non-geometrical ones present.

\end{itemize}

 The plan of the paper is as follows. We devote section
 \ref{geninst} to briefly review
instanton effects in supersymmetric theories and discuss their
stringy analogues. A crucial ingredient is played by anomalous
$U(1)$'s that become massive thanks to a generalization of the
Green Schwarz mechanism in $D=4$. For this to happen a closed
string axionic shift symmetry is gauged. This prevents the
relevant axion from appearing in non-perturbative corrections if
not properly dressed with open string fields charged under the
anomalous $U(1)$. This is  discussed in section
\ref{u1compatib}. In section \ref{z3orbi}, we review basic facts
about the geometry of the underlying orbifold and a schematic
discussion of the quantum stringy corrections to the topological
intersections. We then discuss in section \ref{z3unorient} how to
consistently include open and unoriented strings in the
description. We also review the fate of the anomalous $U(1)$ ('s)
and how discrete and continuous Wilson lines allow to conveniently
break the Chan Paton gauge group. In section \ref{nonperteffz3},
we  specialize our instanton analysis to the case of the ${\bf
Z}_3$ un-orientifold. We discuss non-perturbative superpotentials
induced by both $ED5$'s and $ED1$'s. Then, in section
\ref{fluxedz3} we discuss the superpotential generated by closed
string fluxes, {\it viz.} Scherk-Schwarz shifts (torsion) and R-R
3-form flux included compatibly with the various projections.
Finally, we conclude with some comments on T-duality and the issue
of moduli stabilization.

 {}{}{}{}{}{}{}{}{}{}{}{}{}{}{}{}{}{}{}{}{}{}{}{}{}{}
\section{\label{geninst}General discussion of instanton effects}

Instantons are classical solutions of  the Euclidean field equations
with finite action. Although, strictly speaking, they
represent a set of zero measure in the space of field
configurations, including  quadratic fluctuations and integrating
over the exact moduli may generate tiny, but new and important
effects beyond the reach of perturbation theory. Unfortunately in
pure Yang-Mills or QCD, a reliable computation of these effects is
seriously hampered by IR divergences. The combination of instantons,
holomorphy, anomaly and more recently duality considerations has
proven to be an unprecedented tool in the investigation of
supersymmetric theories. Instantons can generate non-perturbative
corrections to the superpotential in $\cN = 1$ theories
\cite{N1instrev, VYTeffact, ADSsuppot, Seibdual}, thus leading to
the formation of chiral condensates that imply dynamical
supersymmetry breaking in special cases by consideration of the
Konishi anomaly. In $\cN = 2$ theories instantons correct the
analytic prepotential encoded in the periods of an auxiliary
Seiberg-Witten curve \cite{SWN2}. In $\cN = 4$ theories, thanks to
the absence of $R$-symmetry anomalies, they interfere with
perturbation theory and should account for non-perturbative
corrections to anomalous dimensions of unprotected operators
expected on the basis of $S$-duality \cite{N4inst}.

The algebro-geometric construction of  instantons in gauge
theories, that goes under the name of ADHM construction after
Atiyah, Drinfeld, Hitchin and Manin \cite{ADHM}, finds an
intuitive description in open string theory, whereby the gauge
theory is realized on a certain stack of $Dp$-branes and
instantons are represented by a gas $D(p-4)$-branes within the
previous stack \cite{Douglas:1995bn,instfromstr}. ADHM data are
the lowest lying modes of the open strings connecting the
$D(p-4)$-branes with one another or with the $Dp$-branes. In a
supersymmetric setting, these also account for fermionic (zero)
modes. On top of their intrinsic beauty and elegance this kind of
analysis has found a number of applications and proves crucial for
our present purposes.

{}{}{}{}{}{}{}{}{}{}{}{}{}{}{}{}{}{}{}{}{}{}{}{}{}{}
\subsection{Instantons in supersymmetric gauge theories}
\label{gaugeinst}

Pure $\cN =1$ supersymmetric  Yang-Mills theories are expected to
confine and to expose chiral symmetry breaking resulting from the
formation of a chiral condensate for the gaugino. Although the
exact spectrum of bound-states ('superglueballs') is only
approximately known in the strong coupling regime at large $N$
thanks to generalizations of the AdS/CFT correspondence, the
precise value of the chiral condensate in terms of the RG
invariant scale $\Lambda$ can be indirectly derived by means of
instanton calculus exploiting vacuum dominance. Indeed the chiral
correlator \be \langle \lambda\lambda(x_1) ... \lambda\lambda(x_N)
\rangle \ee is dominated by instantons with  instanton number
$K=1$ that give rise to a constant result proportional to
$\Lambda^{3N}$ as expected on supersymmetry grounds and
dimensional analysis. The dynamics of the massive ``glueball"
superfield $S = W^\a W_\a$ is governed by the
Veneziano-Yankielowicz superpotential \cite{VYTeffact} \be
W_{VY}(S) = N S \log{S \over c_N \Lambda^3} \ee where $c_N$ is a
constant depending on $N$ and on the scheme chosen (Strong
Coupling vs Weak Coupling approaches). We will not address this
subtle issue in the present paper.

If one includes matter in the  form of chiral fields, their
classical superfield equations get corrected by the Konishi
anomaly \cite{N1instrev} \be {1\over 4}\bar{D}^2 \Phi^\dagger_I
e^{gV}\Phi^J = {\de W \over \de \Phi^I} \Phi^J + \delta_I{}^J {g^2
\over 32\pi^2} \tr_{\bf R} W^2 \ee This proves useful in setting
the relative strength of the various allowed chiral condensates
or, when this becomes impossible in a supersymmetric vacuum, in
arguing for dynamical supersymmetry breaking. The existence of
flat directions plays a crucial role in this respect. For instance
consider $\cN=1$ SQCD, whereby chiral multiplets in the ${\bf N}$
($Q_i$) and ${\bf N}^*$ ($\tilde{Q}_i$) with $i=1, ... N_f$ are
included. When $N=N_f+1$, instantons generate the  Affleck, Dine,
Seiberg superpotential \cite{ADSsuppot}
\be
W_{ADS} = {\Lambda^{2N+ 1} \over \det(Q \tilde{Q})}
\ee
In the absence of explicit mass
terms this pushes the vacuum to infinity along a flat direction.
If one turns on mass terms of the form
\be W_m = \sum_{ij}m_{ij}Q^i\tilde{Q}^j
\ee
the Konishi anomaly implies
\be
\sum_{ij}m_{ij}\langle Q^i
\tilde{Q}^j \rangle = N_f {g^2 \over 32\pi^2}\langle \lambda \lambda
\rangle
\ee
Since the relevant chiral correlator in this case ($N
= N_f+1$) is
\be {g^2 \over 32\pi^2} \langle \lambda \lambda(x_0)
Q^{i_1} \tilde{Q}^{j_1}(x_1) ... Q^{i_{N_f}}
\tilde{Q}^{j_{N_f}}(x_{N_f})\rangle = \Lambda^{2N+1}~\e^{i_1\cdots i_{N_f}}~\e^{j_1\cdots j_{N_f}}
\ee
one finds
\be
{g^2 \over 32\pi^2} \langle \lambda \lambda\rangle = \Lambda_L^{3}
= (N_f!)^2{\Lambda^{2N + 1} \over \det(Q\tilde Q)}=
\left[{\Lambda^{2N+1}\over N_f^{N_f}}~\det m\right]^{1\over N}
\ee
where $\Lambda_L$ is the RG
invariant scale of the low-energy gauge theory along the flat
direction. Moreover
 \be
\langle Q^i \tilde{Q}^j \rangle =  (m^{-1})^{ij} \Lambda_L^{3} \ee
that clearly shows how the vacuum wanders to infinity when
$\det(m)=0$.

Using decoupling arguments one can generalize the analysis to the
cases $N_f \le N$, whereby the relevant one-instanton dominated
correlator is
\be
\left({g^2 \over 32\pi^2}\right)^{N-N_f} \langle
\lambda \lambda(y_1)... \lambda \lambda(y_{N-N_f}) Q^{i_1}
\tilde{Q}^{j_1}(x_1) ... Q^{i_{N_f}}
\tilde{Q}^{j_{N_f}}(x_{N_f})\rangle = \Lambda^{3N-N_f}~\e^{i_1\cdots i_{N_f}}~\e^{j_1\cdots j_{N_f}}
\ee
where the
exponent  is not unexpectedly the one-loop $\beta$ function
coefficient, $\beta_1 = 3N - N_f$ in this case. In general \be
\beta_1 = 3 \ell({\bf Adj})  - \sum_I \ell({\bf R}_I) \ee where
$\ell({\bf R})$ denotes the Dynkin index of the representation
${\bf R}$, normalized so that $tr_{\bf R}(T^a T^b) = \ell({\bf R})
\delta^{ab}$.

At $N=N_f$ a baryonic branch opens up and  for $[3N/2]> N_f \ge N+1$
SQCD admits an IR free dual `magnetic' description. For $3N > N_f
\ge [3N/2]$  the theory enters the superconformal window. For $N_f>
3N$ the electric theory is trivial / free in the IR.

In more general (chiral) theories, one can use various  symmetry
arguments, including anomalous violation, in order to identify the
relevant one-instanton dominated correlators. In the absence of
flat directions, one can exploit vacuum dominance to extract the
various chiral condensates compatibly with the Konishi anomaly. If
this cannot be satisfied in a supersymmetric vacuum (in which the
LHS vanishes!) one has to infer dynamical supersymmetry breaking.
In the presence of flat directions the vacuum can wander to
infinity in field space.

The  rule of thumb for the one-instanton generation of a
non-perturbative superpotential is the counting of fermionic
zero-modes \cite{N1instrev, VYTeffact, ADSsuppot, Seibdual}. This
number should be two since
\be
L = \int d^2\theta ~W + h.c.
\ee
 In
general there are $ 2\ell_{\bf Adj} $ gaugino  zero modes, \eg
$2N$ for $SU(N)$, and $2\sum_I \ell({\bf R}_I)$ matter fermion
zero-modes. If $\sum_I \ell({\bf R}_I)< \ell_{\bf Adj}$, matter
and gaugino zero-modes are lifted in pairs by Yukawa interactions
\be L_{Yuk} =
\sqrt{2} g ~\phi^\dagger_I \psi^I \lambda
 \ee
In particular for $\ell_{\bf Adj} - \sum_I \ell({\bf R}_I) = 1$,
all matter fermion zero modes are lifted and only two gaugino
zero-modes survive, the ones associated to the broken Poincar\'e
susy. The non-perturbative superpotential acquires the strikingly
simple form \be W_{n-p} =  \Lambda_L^{3} =  {\Lambda^{3\ell_{\bf
Adj} - \sum_I \ell({\bf R}_I)} \over \cH (\Phi)} \ee where
$\cH(\Phi)$ is a chiral gauge invariant, flavor singlet composite
of mass dimension $\Delta_\cH = 2\sum_I \ell({\bf R}_I) =
2(\ell_{\bf Adj} -1)$ and $\Lambda_L$ is the RG invariant scale of
the low-energy gauge theory along the flat direction.

In our stringy application we will need the above  result for
$G=SU(4)\approx SO(6)$ with 3 chiral multiplets in the ${\bf 6}$
dimensional representation, that can be either viewed as the
antisymmetric tensor of $SU(4)$ or as the vector of $SO(6)$. It
easy to see that indeed $\ell_{\bf Adj} - \sum_I \ell({\bf R}_I)= 4 - 3 = 1$ in this case. Actually, as anticipated, an anomalous
$U(1)$ will also play a crucial role in the string setting.
 Another, perhaps more interesting, case would be $G=SU(5)$ with
two chiral multiplets in ${\bf 5} + {\bf 10}^*$. Once again
$\ell_{\bf Adj} - \sum_I \ell({\bf R}_I) = 5 - 2\times( 1/2 + 3/2)
= 1$ and an ADS-like superpotential of the form \be W_{ADS} =
{\Lambda^{11} \over \Phi^2_{\bf 5} \Phi^6_{{\bf 10}^*}} \ee is
generated by instantons. We will not delve into this case any
further although it should admit a $D$-brane realization in string
theory.

 {}{}{}{}{}{}{}{}{}{}{}{}{}{}{}{}{}{}{}{}{}{}{}{}{}{}
\subsection{Instantons in string theory\label{stringinst}}

World-sheet instantons in heterotic and type  II theories have a
long history \cite{wsinst}. They correspond to Euclidean
fundamental string world-sheets wrapping non-trivial cycles of the
compactification space and produce effects that scale as
$e^{-R^2/\ap}$. Depending on the number of supersymmetries and thus
on the number of fermionic zero-modes, they may correct the
two-derivative effective action or they can contribute to
threshold corrections to higher derivative (BPS saturated)
couplings. For type II compactifications on CY threefolds,
preserving $\cN =2$ supersymmetry in $D=4$, world-sheet instantons
correct the special K\"ahler geometry of vector multiplets (type
IIA) or the dual quaternionic geometry of hypermultiplets (type
IIB). Mirror symmetry allows to relate the former to the tree
level exact special K\"ahler geometry of vector multiplets in type
IIB, that can be computed by algebraic methods in terms of the
structure of the so-called chiral ring.  For heterotic
compactifications with standard embedding of the holonomy group
$SU(3)$ in the gauge group, complex structure deformations $U^a$
(with $a=1,..., h_{2,1}$) are governed by the same special
K\"ahler geometry as in type IIB on the same CY threefold, that is
not corrected by world-sheet instantons. Complexified K\"ahler
deformations $T^i$ (with $i=1,..., h_{1,1}$) are governed by the
same special K\"ahler geometry as in type IIA on the same CY
threefold, that is corrected by world-sheet instantons, or
equivalently, as a result of mirror symmetry, by the same special
K\"ahler geometry as in type IIB on the mirror CY threefold with $
\tilde{h}_{2,1}= h_{1,1}$, that is tree level exact. For standard
embedding, the K\"ahler metrics of charged supermultiplets in the
$\bf 27$ ($C^i$ with $i=1,..., h_{2,1}$) and $\bf 27^*$ ($\tilde{C}^a$
with $a=1,..., h_{1,1}$) are simply determined by the ones of the
neutral moduli of the same kind by a rescaling \cite{DKL}. For non
standard embeddings, the situation is not so obvious. In
particular for some time it was believed that the resulting
$\cN=(2,0)$ SCFT would be destabilized by word-sheet instantons.
More recently explicit examples have been constructed where
world-sheet instanton effects conspire to cancel \cite{SilvWit}.

Before turning our attention to $D$-brane instantons,  let us
mention that Euclidean $NS5$-branes wrapping the 6-dimensional
compactification manifold produce non-perturbative effects in
$g_s$ (\ie $e^{-c/g_s^2}$, reflecting the tension of $NS5$-branes)
that qualitatively correspond to 'standard' gauge and
gravitational instantons \cite{BBStrom}.

$D$-brane instantons produce effects that scale as $e^{-c_p/g_s}$,
reflecting the tension of $Dp$-branes \cite{BBStrom}. In type IIB
on CY threefolds, $ED(-1)$, $ED1$-, $ED3$- and $ED5$-brane
instantons, obtained by wrapping holomorphic submanifolds of
complex codimension 3, 2, 1 and 0 respectively, correct the dual
quaternionic geometry in combination with the above mentioned
world-sheet ($EF1$-) and $NS5$-brane ($EN5$-) instantons. In type
IIA $ED2$-instantons (`membrane' instantons) wrapping special
Lagrangian submanifolds, correct the dual quaternionic geometry,
in combination with $NS5$-brane ($EN5$-) instantons. Recall that
the dilaton belongs to the universal hypermultiplet in both cases.

In type I, the presence of $\Omega$9-planes  severely restricts
the possible homologically non trivial instanton configurations.
Essentially  only $ED1$- and $ED5$-branes are topologically stable.
Other (Euclidean) branes can only be associated to instanton with
torsion (K-theory) charges. For other un-orientifolds the
situation is similar and can be deduced by means of T-duality (\eg
for intersecting $D6$-branes one has two different kinds of
$ED2$-branes, for intersecting $D3$- and $D7$- branes one has
$ED(-1)$ and $ED3$-branes, etc).

As mentioned in the introduction there are essentially two extreme
kinds of $ED$-brane instantons. When the $ED$-brane shares the
equivalent of 4 ND directions with a given stack of $D$-branes, it
produces the stringy version of gauge instanton effects
\cite{instfromstr}. This situation is realized when the
$ED$-branes wrap the same cycle as the background $D$-brane and is
point-like in  Euclidean space-time \cite{stringinst}.

 Although we only discuss $D9$ with $ED5$ and $D9$ with
$ED1$, all other cases are essentially related by T-duality to
this one we will focus on. To be precise, in a toroidal orbifold
the most general case is a pair of magnetized $D9$ and magnetized
$ED5$ (the latter is point-like in the space-time directions).
There are conditions that determine which `magnetization' of the
$ED5$ is compatible (due to supersymmetry) with the magnetization
of the `background' $D9$. An index theorem also fixes the number
of zero modes and, as a result, determines whether the magnetized
$ED5$ may produce a non zero F-term.

On the opposite side, when the $ED$-brane shares the equivalent of
8 ND directions with a given stack of $D$-branes, it produces
`new' genuine stringy instanton effects that cannot be reproduced
by standard gauge instantons \cite{stringinst}. In particular, in
the Type I case, $ED5$-branes qualitatively behave as gauge
instantons for $D9$-branes but as non-standard ('octonionic'?)
instantons for $D5$-branes (if present). On the other hand
$ED1$-branes qualitatively behave as gauge instantons for
$D5$-branes wrapping the same internal cycles but as non-standard
 instantons for $D9$-branes or for $D5$-branes
wrapping orthogonal cycles.  We will consider precisely the effect
of $ED1$ on $D9$'s later on.

When world-volume
(magnetic) fluxes are turned on a given stack of $D$-branes or a
$ED$-branes the resulting effect is intermediate. We will not
address this very interesting issue in the present paper since we
will only work with isotropic (not {\it coisotropic}!) $D$ and $ED$-branes.

As shown in \cite{instfromstr}, $ED(p-4)$-branes  within $Dp$-branes
precisely reproduce the instanton action, the ADHM data and as a
result the instanton profile together with the associated
zero-modes. One can thus proceed along the lines of the field
theory analysis, \ie identify the relevant one- or $K$- instanton
dominated amplitude and infer the form of the non-perturbative
correction to the effective action.

The ADHM data correspond to strings connecting  $ED(p-4)$-branes
with one another or with $Dp$-branes. We denote by $N$ the number of $Dp$ branes
and by $K$ the respective number of $ED(p-4)$. The corresponding bosonic
vertex operators for $ED(p-4)$-$ED(p-4)$ strings ($K\times K$ of them)
are of the form
\be V_a = a_\mu
e^{-\varphi}\psi^\mu T_{K\times K} \ee for the non dynamical gauge bosons,
where $\mu$ denote the D-D
space-time directions with no momentum ($p=0$), and
\be V_\chi = \chi_i
e^{-\varphi}\psi^i T_{K\times K}
\ee
for the non dynamical transverse scalars, where $i$ denote the internal
directions longitudinal (NN) or orthogonal (DD) to the $ED$-branes
not affected by twist / orbifold projections (if any).

The low-lying $ED(p-4)$-$Dp$ strings ($K\times N$ of them plus
conjugate) admit bosonic vertex operators of the form \be V_w =
\sqrt{g_s\over v_{p-3}} w_\alpha e^{-\varphi}\prod_\mu
\sigma_{\mu} S^\alpha T_{K\times N} \ee where $\sigma$ are $Z_2$
twist fields and $S^\alpha$ is a spin field of a given chirality
(left) along the 4 ND directions and  the overall normalization,
suggested in \cite{instfromstr} and then used in \cite{stringinst}
is crucial in order to obtain the correct field theory limit.

The instanton action coincides with ($K$ times) the gauge  kinetic
function since the $ED(p-4)$-branes under consideration wrap
exactly the same (supersymmetric) cycle as the $Dp$-branes. As
such \be S_{inst, A} = f_A(S,T,U,Z,\Phi) \ee can depend on moduli
of various kinds: dilaton ($S$), K\"ahler ($T$), complex structure
($U$), twisted ($Z$), open string ($\Phi$) both charged and
neutral. For $D9$ branes in orbifolds, for instance \be f_{D9} = S
+ B_I Z^I + \Delta (T,U) + ... \ee where $B_I$ denote the disk
tadpole of $Z_I$ (twisted moduli) and $\Delta$ denote one-loop
threshold corrections, which turn out to be a constant for the
case of the ${\bf Z}_3$-orbifold we are interested in here.

By computing disk amplitudes with insertions  of $V_a$, $V_\chi$,
$V_w$, $V_{w^\dagger}$ and their superpartners, one can
reconstruct the classical profiles needed to compute
non-perturbative contribution to scattering amplitudes (see \cite{instfromstr}).

The analysis of the other kind of instantons is different
\cite{stringinst}. The prototype is the $D9$, $D1$ system whose (multi)-instanton configuration was first analyzed
in \cite{bfkov}. The
number of N-D directions is 8 in this case and the lowest lying
modes of the open string stretched between the $N$ $D9$'s and the
$K$ $D1$'s are massless fermions with a given chirality along the
common two NN dimensions. In the standard case of type I strings
there are 32 such chiral fermions ($\lambda^A$) that precisely
reproduce the gauge degrees of freedom of the `dual' heterotic
string. In addition from the 1-1 sector there
are 8 transverse bosons $X^I$ in the $8_v$ of the $SO(8)$
R-symmetry group and as many Green-Schwarz type fermions $S^a$ of
opposite chirality (say Left) in the $8_s$ giving rise to an $\cN
= (8,0)$ theory on the $D1$ world-sheet. The 32 massless
right-moving $\lambda^A$ are inert under the 8 left-moving susy
$Q_{\dot{a}}$ in the $8_c$.

After compactification to $D=4$ on a manifold with non-trivial
holonomy some of the global supersymmetries are broken and the
corresponding $D1$ world-sheet theory changes accordingly. In
particular the left-moving degrees of freedom include the
surviving superspace variables $\Theta$ and $\bar\Theta$
descending from the GS fermions\footnote{For an instanton contribution to the superpotential
the $\bar\Theta$ should be either massive or projected out. We will confirm this in section
\ref{wed1}.}
\be
V_\Theta = \Theta_\a S^\alpha
\Sigma_{+3/2} e^{-\varphi/2} \quad ,
\ee
where $S^\alpha$ is a
dimension 1/4 spin field in the space-time directions and $\Sigma$
is a dimension 3/8 internal spin field. Four surviving
(non-dynamical, no $p$) massless bosons, \be V_a = a_\mu
e^{-\varphi} \psi^\mu \ee corresponding to the motion along the
four flat space-time directions, are always present. Extra
massless bosons, corresponding to the motion along the internal
directions, may appear depending on the possibility of deforming
/sliding the cycle wrapped by the brane, however such cycles will not contribute to the superpotential.
Rigid cycles would admit
no such motions. The number of massless chiral fermions $\lambda$
\be V_\lambda = \sqrt{g_s} \lambda_R e^{-\varphi/2} S^{-}
\prod_\mu \sigma_{(\mu)} \prod_I \sigma_{(I)} \ee where $S^{-}$ is
a dimension 1/8 right-handed spin field along the two NN
directions, and $\sigma_{(\mu)}$ and $\sigma_{(I)}$ are $Z_2$
twist fields along the four space-time and as many internal ND
directions, depends on the number of $D9$'s of a given type in the
vacuum configuration that `intersect' the $D1$'s. {\it Mutatis
mutandis} one can identify the relevant degrees of freedom for the
other cases (\eg $ED3$ in a background of $D3$, $ED2$ in a
background of intersecting $D6$'s, etc) \cite{stringinst}.

As described in \cite{instfromstr}, one has to integrate over the
`non-dynamical' modes living on the world-volumes of the $EDp$
under consideration. As a result one can generate non-perturbative
corrections to the (super)potential. A comment however is in
order. These effect are non-perturbative in that they scale as
$e^{-T_{EDp} V_{EDp}}$. Since $T_{EDp}\approx 1/g_s (\ap)^{p+1/2}$
these effects are non-perturbative in $g_s$. Yet they {\it a
priori} depend on different moduli (through the dependence of
$V_{EDp}$ on various $Z$'s) from the ones that appear in the gauge
kinetic function(s) so they cannot in general be identified with
`standard instantons'. In fact one can envisage the possibility of
turning on magnetic fluxes on the world-volume of the $EDp$ that
allow one to interpolate\footnote{This interpolation is discontinuous due to the quantization of the
magnetic fluxes but may become quasi-continuous at large volumes.} between one kind
of $EDp$ (\eg a `standard instanton') \cite{instfromstr} and a
different kind of $EDp$ (\eg a new stringy instanton)
\cite{stringinst}.

As shown in \cite{stringinst}, elaborating on the $g_s$ power
counting introduced in \cite{instfromstr}, the relevant diagrams
are disks with insertions of the non-dynamical vertex operators
$V_\Theta$ (connecting $EDp$'s with themselves) and $V_\lambda$
(connecting $EDp$'s with the background $Dp'$) with or without
insertions of the dynamical vertex operators $V_A$ etc
corresponding to the massless excitations of the vacuum
configuration of (intersecting / magnetized ) unoriented
$D$-branes. Disks without insertions of the latter type yield
after exponentiation the `instanton action' (including
interactions of the pseudo zero-modes $\lambda$'s etc). Disks with
one dynamical vertex produce the classical profile that is needed
for the computation of the non-perturbative amplitudes. Disks with
more insertions contribute to higher-order corrections that can be
neglected at first and consistently incorporated later on by
symmetry arguments. One loop diagrams with no insertions
should produce subtle numerical prefactors that can conspire so as
to cancel a given type of non-perturbative F-terms
\cite{SilvWit}.

For a supersymmetric instanton there are two $\Theta$ zero-modes.
Suppose that one also has $2n$ $\lambda$ zero-modes. Then one can
compute an F-term either by a combination of $n$ disks, each with
two $\lambda$ insertions, out of which either $n-2$ of them
accommodate one $V_\phi$ and two of them one $V_\psi$ each, or
$n-1$ of them accommodate one $V_\phi$ and two of them one $V_F$
(vertex operators for auxiliary fields can be written in a non BRS
invariant form that give anyway sensible results). Integrating
over $\Theta$'s and $\lambda$'s yield a superpotential term of the
form \be W = e^{-T_{EDp} V_{EDp}(Z)} \Phi^n \ee where the notation
is schematic in that $\Phi^n$ denotes a gauge invariant monomial
of degree $n$ in the superfields $\Phi^i = \phi^i + \Theta \psi^i
+ \Theta^2 F^i$ and $Z$ denote moduli fields whose dependence is
tightly constrained by geometric and other symmetry
considerations.

 {}{}{}{}{}{}{}{}{}{}{}{}{}{}{}{}{}{}{}{}{}{}{}{}{}{}
\subsection{Compatibility of bulk isometries and
non-perturbative effects}
\label{u1compatib}

On general grounds a chiral field $Z$ whose pseudoscalar axionic
components $\zeta = ImZ$ shifts under some local anomalous $U(1)$
cannot appear as such in a (super)potential term. However it can
appear dressed with other chiral fields that are charged under the
$U(1)$. $U(1)$ invariance puts stringent constraints on the form
of the possible superpotential terms. Since the axionic shift is
gauged it must be a symmetry of the kinetic term. This is only
possible when no non-perturbative (world-sheet or $D$-brane
instanton) correction spoils the tree level (in fact perturbative)
PQ symmetry of the K\"ahler potential. In turn this means that the
gauging procedure corresponds to turning on fluxes such that the
potential instanton corrections in $Z$ are in fact disallowed. In
practice, this means the corresponding wrapped brane is either
anomalous (\`a la Freed Witten) \cite{FreedWit} or destabilized
due to the flux \cite{KasPurTom}. In this respect chiral fields
that appear in superpotential terms tend to have `quasi-canonical'
kinetic terms compatible with their continuous shift symmetries
that are gauged.

This state of affairs has been checked in various cases
\cite{mmmm}(\eg $ED3$ in flux compactifications with $D3$ and
$D7$'s) and will also hold  in the case we are going consider \ie
the ${\bf Z}_3$ orbifold with Wilson line breaking $U(12)\times
SO(8)$ to $U(4)\times G_{H}$ with $G_{H}$ a hidden gauge group
such as $U(4)_{CFT}^3$ or $U(4)_{{\cal N}=4}$ or $U(1)^{4}$.

In general, the Bianchi identities for the `total' R-R field
strength $G = \sum_p G_p$ is encoded in \be
 D G = \Pi[branes] \wedge e^F
 \ee
where $\Pi[branes] = \sum_p \Pi_{9-p}[Dp]$ denotes a formal sum of
$(9-p)$-forms along the directions orthogonal to the branes
present in the background, $D = d + \cT + H$ where $\cT$ is the
geometric torsion \`a la Scherk-Schwarz and $H$ is the NS-NS
3-form field strength.\footnote{We neglect from this discussion the presence of the curvature terms in the WZ action as well as the non-geometric fluxes
${\cal Q}$ and ${\cal R}$ for simplicity. The relevant non-geometric fluxes will be restored later on.}
 For the lowest type IIB  R-R forms one gets
\be dG_1 +\cT \circ G_1= F_2 \wedge\Pi_0(D9) + \Pi_2(D7)
\label{1}
\ee
 and
\be
dG_3 + \cT \circ G_3 + H_3 \wedge G_1 = F_2\wedge F_2 \wedge\Pi_0(D9) +
F_2 \wedge\Pi_2(D7) + \Pi_4(D5)
\ee
and so on, were we denoted the action of the geometric
torsion on forms as
\be
\cT\circ A_p\equiv (p+1){\cT^s}_{[\m_1\m_2}A_{\m_3\cdots s\cdots m_{p+1}]}
\ee

Barring torsion ($\cT = H = 0$) and D7-branes (\ref{1}) yields
\be
dG_1 = F_2\;\;.
\ee
Integrating this on the closed world-volume of
a D-string we obtain
\be
\int F_2 = \int dG_1 = 0
\ee
In general the
axionic shifts are given by
\be
\delta \beta^I_{R-R}(x) =
\alpha^a(x) \int_{\cC_I} tr(F_a)
\ee
where $\cC_I$ represents a
basis of 2-cycles dual to the harmonic 2-forms $\omega_I(y)$ that
appear in the expansion of the R-R 2-form
\be
C_{R-R} (x,y) =
\beta^I_{R-R}(x) \omega_I(y) + ... (massive)
\ee
This means one
cannot wrap an $ED$-string on any cycle $\cC$ such that
\be
\int_{\cC} tr(F_a) \neq 0
\ee
\ie around the cycle dual to the R-R
axion whose shift symmetry is gauged. This remains true even if
$G_1$ and $F_2$ are odd under $\Omega$ (worldsheet parity, \ie in
unoriented theories with $D9$ and $D5$) very much as in the
`standard' construction with $D3$ and $D7$ branes the presence of
$H_3$ and $F_3$ fluxes ($``G_3"$ fluxes) obstructs some
$ED3$-brane instantons even if the fluxes are odd under  $\Omega'=
\Omega(-)^{F_L}\sigma$. For the ${\bf Z}_3$ orbifold we will
momentarily see that $\cC$ is a democratic linear combination of
the 27 twisted cycles corresponding to collapsed $P^2$s at the
orbifold points.

\section{Review of type II ${\bf Z}_3$ orbifold}
\label{z3orbi}

To construct the ${\bf Z}_3$ orbifold we act on the three complex
coordinates of the ${\bf T}^6$ torus, $z^I$, $I=1,2,3$ as \be z_I
\rightarrow \omega z_I\sp \omega = e^{2\pi i/3} \ee To be a
symmetry of ${\bf T}^6$, we must constraint the metric and the
NS-NS two-form to be \be ds^2=G_{I\bar I}~dz^Id\bar z^{\bar I}\sp
B_2=B_{I\bar I}~dz^I \wedge d\bar z^{\bar I} \ee The $3\times 3$
complex matrix $G_{I\bar I}$ is hermitian while $B_{I\bar I}$ is
anti-hermitian. The ${\bf Z}_3$ action is chosen so that the
holomorphic three-form $dz^1 \wedge dz^2 \wedge dz^3 $ is
invariant. Therefore the ${\bf Z}_3$ orbifold is a (singular) CY
three-fold.

Compatibility with the ${\bf Z}_3$ projection
freezes out all the complex structure deformations but allows 9
untwisted deformations of the K\"ahler structure $dz^I \wedge
d\bar{z}^{\bar{J}}$, so that $h_{1,1}^{untw}=9$, while
$h_{2,1}^{untw}=0$. The complex untwisted K\"ahler moduli are $G_{I\bar J}+
B_{I\bar J}$.
They can both be expanded in the standard basis of hermitian matrices
 ${\cH}_{ij}$ as
\be G=t_{ij}{\cH}^{ij}\sp B=ib_{ij}\cH^{ij} \ee The resulting
moduli space of untwisted complexified K\"ahler moduli is \be
\cM^{untw}_{(1,1)} = {SU(3,3) \over SU(3)\times SU(3)\times U(1)}
 \ee
It is a special K\"ahler manifold with (holomorphic) prepotential
\be \cF_{unt}=\det[T]={1\over
3!}\e^{I_1I_2I_3}\e^{J_1J_2J_3}X_{I_1J_1}X_{I_2J_2} X_{I_3J_3}\sp
X_{IJ}=t_{ij}+ib_{ij} \ee The associated K\"ahler potential is
given by the special geometry formula
\be
K_{unt}=-\log\left[2(\cF_{unt}+\bar
\cF_{unt})-\sum_{IJ}(X_{IJ}-\bar X_{IJ})\left({\partial \cF_{unt}\over
\partial X_{IJ}}+
{\partial \bar \cF_{unt}\over \partial \bar X_{IJ}}\right)\right]
\ee
$$
=-\log[\det[Re[X]]]
$$
where the $X_{IJ}$ are
the inhomogeneous K\"ahler coordinates. The two-forms dual to the
moduli $X_{IJ}$ are $\omega_{IJ}=dz^I\wedge d\bar z^J$ in
one-to-one correspondence with the non-trivial two-cycles of the
torus, with intersection form \be \int
\omega_{I_1J_1}\wedge\omega_{I_2J_2}\wedge\omega_{I_3J_3}=\e_{I_1I_2I_3}
\e_{J_1J_2J_3} \ee The associated four-forms are
$\omega^{IJ}=\e^{II_1I_2}\e^{JJ_1J_2}dz^{I_1} \wedge
dz^{I_2}\wedge d\bar z^{J_1}\wedge d\bar z^{J_2}$.

${\bf Z}_3$ has 27 fixed points corresponding
to as many 'exceptional divisors', $E_{i}$, $i=1,2,\cdots,27$.
They are codimension-one complex
submanifolds which are homologically non trivial.
There are as many
twisted (1,1)-forms, so that $h_{1,1}^{twist}=27$, while
$h_{2,1}^{twist}=0$.
There are three fixed points per two-plane so that we will label the 27 fixed
points by $f_{i}$.

The orbifold is resolved by excising a small neighborhood around
the fixed points and gluing in ${\bf Z}_3$ Eguchi-Hanson-like
balls each with Euler number $\chi=3$. Since the original torus
has $\chi=0$ and each excised point has $\chi=1$ we obtain the
total Euler number \be \chi({\bf T}^6/{\bf Z}_3)={0-27\over
3}+27\cdot 3=72 \ee The two-forms dual to the exceptional cycles
$\omega^{i}$ have the non-trivial intersection \be \int
\omega^{i}\wedge\omega^{i}\wedge\omega^{i}=1\sp \forall ~~i \ee
while all other intersections between different fixed points, or
with the untwisted ones, vanish. We will denote the 27 associated
complex twisted moduli by $Z_{i}$.

The parent type IIB theory enjoys local $\cN = 2$   supersymmetry.
In addition to the supergravity multiplet and the universal
hypermultiplet, whose four scalars can be identified with the
dilaton, the R-R axion, the NS-NS 2-form and the R-R 2-form both
dual to axions, the massless spectrum contains $36 =h_{1,1}^{untw} +
h_{1,1}^{twist} $ hypermultiplets. Their scalar
components are the K\"ahler deformations of the metric which are
tri-complexified by the NS-NS 2-form and the R-R 2-form and the
self-dual 4-form.

The unoriented projection preserves local $\cN = 1$ supersymmetry,
thus eliminating the R-R graviphoton and one (linear combination
of the) gravitini. Each (linear/) hypermultiplet produces a
(linear/) chiral multiplet. The standard $\Omega$ projection
retains the dilaton, the metric and the R-R 2-form, but one can
envisage different (not necessarily) equivalent truncations. For
instance, after 6 T-dualities one ends up with an essentially
equivalent theory with $\Omega 3$-planes, that retains the
dilaton, the R-R axion, the metric and the R-R 4-form.

A relative of the ${\bf Z}_3$ orbifold is obtained by  acting with
another ${\bf Z}_3$ transformation that rotates the coordinates
$z^1,z^2$ as $z^i\to \omega z^i$. There are two options here, the
free action or the non-free action. In both cases out of the 9
untwisted moduli $X_{IJ}$ only three survive the new projection:
$T_{11},T_{22},T_{33}$. In the twisted sector the situation
depends on the action. Most interesting for us will be the free
action where the the extra ${\bf Z}_3$ transformation is
accompanied by a ${\bf Z}_3$ translation on the third torus along
the lattice. Because the old fixed points remain fixed, the new
orbifold has the same twisted sector as before, and the same
number of massless twisted moduli. On the other hand the non-free
${\bf Z}_3\times {\bf Z}_3'$ orbifold has 81 fixed points.

In the free case, the associated prepotentials as well as K\"ahler
potentials can be obtained directly by restriction from those of
the ${\bf Z}_3$ orbifold studied above. In this last case, the
string instanton corrections to the prepotential have been studied
in \cite{candelas}.

\section{The ${\bf Z}_3$ orientifolds with Wilson lines}
\label{z3unorient}

The standard $\Omega$-projection of the closed string spectrum
generates R-R tadpoles that can be canceled by introducing
$D9$-branes and their (unoriented) open string excitations.
Denoting by $\gamma_{{\bf Z}_3}$ the projective embedding of the
orbifold group in the Chan-Paton group, twisted tadpole
cancellation requires $Tr (\gamma_{{\bf Z}_3}) = - 4$ in addition
to the `standard' untwisted tadpole condition $Tr ({\bf 1}) = 32$
\cite{chiras}. Imposing $\gamma_{{\bf Z}_3}^3= 1$ and
$\gamma_{{\bf Z}_3}^\dagger = \gamma_{{\bf Z}_3}^{-1}$, allows to
set
 $\gamma_{{\bf Z}_3} = ({\bf 1}_{_{N\times N}},
\omega {\bf 1}_{_{M\times M}}, \bar\omega{\bf 1}_{_{\bar M\times
\bar M}})$ so that \be N + M + \bar M = 32 \qquad N + \omega M +
\bar\omega \bar M = -4  \qquad M = \bar M \ee yielding $N=8$, $M =
\bar M = 12$. Due to  the $\Omega$ projection the resulting gauge
group is $SO(8)\times U(12)$. In addition one has three
generations of $({\bf 8},{\bf 12})_{+1}$ plus  $({\bf 1},{\bf
66}^*)_{-2}$, resulting from the breaking of $\cN =4$ SYM to
$\cN=1$ SYM plus three chiral multiplets, all transforming the
same way under ${\bf Z}_3$. The $U(1)$ is anomalous, \ie $t_3 \neq
0$ where the mixed anomaly trace is \be t_3 \equiv Tr[Q_f T^aT^a]
= \sum_f Q_f \ell ({\bf R}_f) \ee with $f$ running over chiral (L)
fermions with charge $Q_f$ and $\ell ({\bf R}_f)$ is the Dynkin
index of the representation ${\bf R}_f$ of the non-abelian group
($G=SO(8)\times U(12)$) the fermions belong to. The generalized GS
mechanism entails a mixing between $V$, the U(1) vector
superfield, and a 'democratic' combination of all twisted chiral
multiplets. Indeed, if we define $Z=\sum_{i=1}^{27}Z_i$ then \be Z
+ \bar Z \rightarrow Z + \bar Z - M V \ee where $M \approx M_s$ is
a mass parameter (computed in \cite{art}), so that under \be V
\rightarrow V+i\alpha - i \bar\alpha \ee one has \be Z \rightarrow
Z+i M \alpha \ee

Indicating the gauge kinetic functions of the non abelian gauge groups by
\be
f_a(T,S;Z; C, A) = f_a(T,S) + C_a Z + .... \label{anomaly2}
\ee
anomaly
cancellation requires
\be
M_a C_a = t_{3,a}
\label{anomaly1}\ee \cite{ib}.

Chiral multiplets $C^{I r}_i$ in the $({\bf 8},{\bf 12})_{+1}$ and
$A^I_{[rs]}$ in the $({\bf 1},{\bf 66}^*)_{-2}$  interact via the
tree level superpotential \cite{chiras}\be W(C,A; T,S; Z) = {1
\over 3!2!} Y(T,S;Z) \epsilon_{IJK} \delta^{ij} C^{I r}_i C^{J
s}_j A^K_{[rs]} \quad . \ee In the T-dual descriptions in terms of
$D3$-branes, when all the branes are at the same fixed point,
$Y(T,S,Z)$ should only depend on the overall volume multiplied by
$e^{-\phi}$. However if regular $D3$-branes move into the bulk,
there could be open string instanton contributions too between far
away branes, as can be checked by explicit computation of a disk
diagram. Dependence on the closed string (un)twisted moduli is
highly constrained by the axionic (shift) symmetries of the axions
contained in $T,S$ and $Z$ (we have collectively labeled by $T$
the untwisted moduli and $Z$ the twisted ones).

Consistently with the above picture, non-perturbative $F/D$-string instanton
corrections are allowed in this case but no perturbative correction that would
spoil the universal  axion ($Im S$) PQ symmetry.

Clearly the physical Yukawas are renormalized as a consequence of
the renormalization of the K\"ahler potential. Higher order terms
in the neutral combination $C C A$ can appear however. Yet terms
containing $Pfaff(A)$ have a nontrivial $U(1)$ (anomalous) charge
and can only be produced non-perturbatively, if at all, because
there is no way to
  contract the indices in a cyclic fashion to produce
$\epsilon^{r_1 ... r_{12}}$ \cite{Niretal,bere1,adks}.
$U(1)$ symmetry prevents them from
appearing in perturbation theory even from non-planar graphs. The
situation may change if one takes into account the non-trivial
U(1) gauge transformation properties of $Z$.

In order to study this possibility, it is convenient to turn on
(`continuous') Wilson lines $\gamma_W$ along the flat directions
of the potential and generically break $SO(8)\times U(12)$ to
$U(4)_{fp} \times U(1)^4$ \cite{torwils, cvetwils}. Special
(`discrete') choices of the Wilson lines correspond to symmetry
enhancement \cite{opensys}. For instance $\gamma_W$ commuting with
$\gamma_{{\bf Z}_3}$ breaks $SO(8)\times U(12)$ to $SO(8-2n)\times
U(12-2n)\times U(n)^3$. In particular for $n=4$ one gets
$U(4)_{fp}\times U(4)^3$ coupled to 3 generations of chiral matter
in the $({\bf 6}_{-2};{\bf 1}_0, {\bf 1}_0,{\bf 1}_0 )$ plus
$({\bf 1}_{0};{\bf 4}_{+1}, {\bf 4^*}_{-1},{\bf 1}_0) $, $({\bf
1}_{0};{\bf 1}_0, {\bf 4}_{+1}, {\bf 4^*}_{-1}) $, $({\bf
1}_{0};{\bf 4^*}_{-1},{\bf 1}_0,{\bf 4}_{+1} ) $. Notice that the
$U(4)^3$ part correspond to a decoupled conformal theory living on
a stack of 4 regular branes \ie 4 branes together with their 6
images under ${\bf Z}_3$ and $\Omega$. In the T-dual description
in terms of $D3$-branes, the latter are located at  two fixed
points different from the origin which are mapped into one another
by $\Omega$. One can further break $U(4)^3$ to $U(4)_{diag}$ with
3 adjoint chiral multiplet thus reconstructing the field content
of $\cN=4$ SYM.
 Finally by turning on VEV's for the six adjoint scalars
 one generically breaks the group to $U(1)^4$.
 One can turn on internal magnetic fields along these
 three $U(1)$'s.
 We refrain from doing so here.

The case of the ${\bf Z}_3\times {\bf Z}_3'$ orbifold where  the
second ${\bf Z}_3$ action is free is similar. Indeed as long as
the size of the Scherk-Schwarz one-cycle is non-zero, the extra
${\bf Z}_3$ acts as a simple projection in the low energy sector.
Moreover, it does not induce additional tadpoles and therefore the
open sector is similar to the ${\bf Z}_3$ one, module the overall
extra ${\bf Z}_3$ projection \cite{ads}.

\section{Non-perturbative effects in the ${\bf Z}_3$ orientifold}
\label{nonperteffz3}

We are ready to discuss non-perturbative effects in the ${\bf
Z}_3$ orientifold. We will start with the effect induced by
wrapped Euclidean D5-branes ($ED5$-branes) that are expected to
reproduce gauge instanton effects. We then consider the effects
due to wrapped Euclidean D1-branes $ED$-strings.

\subsection{Wrapped Euclidean D5-branes}

At the point where $G=U(4)_{fp}\times U(4)_{diag}$, with 3 ${\bf
6}_{-2}$ for the first factor, instanton calculus is reliable.
This is due to the fact that along the flat directions the gauge
group is broken in such a way that no light charged matter
survives. Indeed along the flat directions where the group is
broken to $G_L = SO(3)$, instantons in the resulting pure ${\cal
N}=1 $ theory induce gaugino condensation with $W = \Lambda_L^3$.
The matching condition  between $\Lambda_L$ and $\Lambda$, the RG
invariant scales of the low and high energy theories respectively,
allows one to identify this superpotential with
\be W =
\Lambda_L^3 = {\Lambda^9 \over \det_{I,J}(\delta^{ab}A^I_a A^J_b)}
\ee
 where $A^I_a ={1\over 2}\Gamma_a^{rs} A^I_{rs}$ with
$a=1,...,6$
 are the three chiral multiplets in the ${\bf 6}$ of $SU(4)$,
and $\Gamma_a^{rs}$ are
Weyl blocks of 6-d Dirac matrices.
 In general the argument applies whenever
 $\ell_A - \sum_C \ell_C = 1$, where $A$ denotes the
 adjoint representation and $C$ runs over chiral multiplets.
 Indeed,  in our case $\ell_A = 4$
 and $\sum_C \ell_C = 3 $.

In string theory one expects (up to an overall numerical factor)
\be
W(S,T,Z; A) = {e^{f(S,T,Z)}\over \cH(A)}
\ee
where
\be
f(S,T,Z) = f_{tree}(S,Z) + f_{1-loop}(T,Z)
\ee
is the gauge
kinetic function with\footnote{A dependence of the holomorphic  gauge kinetic  function
on the open string moduli $A$ would entail,
for reasons of U(1) gauge invariance, a further exponential dependence on $Z$. This would imply ``instanton corrections for instanton corrections".
We deem that such dependence is unlikely.}
\be
f_{tree}(S,Z) = S + C Z
\ee
while
\be
f_{1-loop}(T,Z=0) = f_1
\ee
a constant  independent of $T$'s as
originally observed in \cite{ABD} and confirmed in \cite{ABSS}
following previous work on heterotic orbifolds \cite{DKL}. The $Z$
dependence is harder to determine. $S$ independence is
from loop counting. One can indeed check that under simultaneous
$U(1)$ transformations of the $A$'s and shift of $Z$, the
superpotential $W$ is invariant. Indeed based on the mixed
$U(1)\times SU(4)^2$
anomaly \be t_{144} = -2 \times 3 \times 2 =
-12 \ee
 one deduces that $Z$ must shift as
\be
Z \rightarrow Z - {12\over C} i \alpha
\ee
as shown in (\ref{anomaly2},\ref{anomaly1}).
This is
exactly what is needed to cancel the transformation of the
denominator generated by
\be
 A \rightarrow e^{-2 i \alpha} A \quad
.\ee

As previously stated in general terms, the instanton action is
given by the world-volume  of an $ED5$ wrapping the entire
orbifold and this is exactly given by the gauge kinetic function
of the $D9$ branes, including the shift $CZ$. Notice that in the
present case, generation of a non-vanishing AdS-like
superpotential heavily relies on the presence of the
doubly-charged anti-symmetric representations (${\bf 6}_{-2}$) of
the $U(4)$ Chan-Paton group.

In a recent paper \cite{lustADSsuperpot}, a detailed stringy
derivation of the ADS superpotential has been given for the case
of SQCD with gauge group $SU(N_c)$ with $N_f=N_c - 1$ massless
flavours or $Sp(2N_c)$ with $2N_f = 2N_c$ flavours. The case
$SO(2N_c)$ with $N_f = N_c$, we focussed on above for $N_c = 6$,
was only touched upon. In the construction of
\cite{lustADSsuperpot} the gauge theory is realized on a stack of
D6-branes and the flavour symmetry is generated by another stack
of D6-branes intersecting the previous one in a non chiral
fashion. The relevant instanton is an ED2 wrapping the same cycle
as the stack of $N_c$ D6-branes. By careful integrating the
supermoduli the precise form of the  ADS superpotential was
reproduced in the field theory limit. In our case the gauge theory
is realized with branes at a singularity and we have determined
the form of the ADS-like superpotential by holomorphy, dimensional
analysis, $U(1)$ anomaly (fermion zero-mode counting) and flavour
symmetry. We leave it as an open problem to derive the ADS-like
superpotential directly from a full-fledged string computation
along the lines of \cite{lustADSsuperpot}. In order to do so one
has to properly integrate the supermoduli that comprise massless
strings stretching from ED5 to the D9's and those of the ED5
itself that should support a $U(1)$ Chan-Paton group that should
enhance to $Sp(2)$.

\subsection{Wrapped D-strings\label{wed1}}

We would now like to discuss non-perturbative effects induced by
wrapping $ED1$ around topologically non-trivial two-cycles.
This configuration was first considered in \cite{bfkov}, where the ED1 instanton corrections to the $F^4$ and $R^4$
couplings were computed in the toroidally compactified type-I theory. These corrections were originally
obtained by heterotic/type I duality
from the one-loop string instanton corrections in the heterotic string but subsequently justified from the ED1 instanton  point of view.
This computation also gave  a detailed account of the multi-instanton contributions and their subtleties.

In our case we will first show  that unlike the $\Theta$ zero-modes
present and described in section \ref{stringinst}, the $\bar\Theta$ zero modes are absent.
We start with the supersymmetry  in D=10 for $D_9$-branes: the 16 of
SO(10).
Adding ED1's in flat space-time
we decompose $16 \to 8_s^{+1/2}  + 8_c^{-1/2}$ under  $SO(8)\times SO(2)$.
One of the two spinors is projected out so we assume that  $8_c^{-1/2}$
remains. It generates ${\cal N}=(8,0)$ world-sheet supersymmetry with respect to which
which the fermions $\lambda$, in the D9-ED1 sector are inert.
After compactification on the orbifold,
SO(8) is broken at least to $SO(4)_{Min}\times SO(4)_{int}$ where $SO(4)_{Min}$ is the Lorentz symmetry of flat space-time
Therefore,
$8_c \to (2_L, 2_L) + (2_R, 2_R)$.
In order to have a surviving supersymmetry the orbifold projection $g = \exp (i
w_i J_i)$ must be such that
$w_1 + w_2 + w_3 = 0$.
The surviving spinors are the lowest components of $(2_L)_{int}$ i.e. $(-1/2,
-1/2)$ which when combined with the -1/2 helicity with respect to the  the 'world-sheet' SO(2)
yield $-1/2 w_1  -1/2 w_2 - 1/2 w_3 = 0$.
Clearly choosing a different projection with
$\pm w_1 \pm w_2 \pm w_3 = 0$
a different but unique internal spinor component will survive.
As a result only one $SO(4)_{Min}$ chirality of the supersymmetry survives i.e.
$(2_L; (-1/2, -1/2); -1/2)$ of $SO(4)\times SO(4)\times SO(2)$.
Obviously the second SO(4) is not a symmetry, but it is  helpful
in the above decomposition.
In a smooth CY the four supersymmetry charges are
$Q_\alpha = S_\alpha \eta$, $
\bar{Q}_{\dot\alpha} = C_{\dot\alpha}\eta^\dagger$
where $\eta$ and $\eta^\dagger$ are the two covariantly constant spinors of
opposite SO(6) chirality (or U(1) charge, under $SO(6)\to SU(3)\times U(1)$). Only one
of the two has the correct chirality under the SO(2) of the ED1
world-sheet.

In order to determine which kind of superpotential term is
generated one has to count the number of fermionic zero modes
$\lambda$'s stretching from the $ED1$ to the background $D9$'s.
Depending of the 2-cycle $\cC$ wrapped by the $ED1$,
\ie on the
restriction $V\vert_\cC$ of the vacuum gauge bundle $V$ to $\cC$,
$\lambda$'s transforming in the ${\bf 4}_{+1}$ of $U(4)_{D9}$ may
appear. These can couple at the disk level with the scalar
component $a^I$ of the multiplet $A^I$ in the ${\bf 6}_{-2}$. Let
us indicate this coupling by \be L = m_I(\cC) A^I_{[rs]}
\lambda^r_\cC \lambda^s_\cC \ee where $m_I(\cC)$ depends on the
cycle $\cC$ wrapped by the $ED1$\footnote{A formal expression for $m_I(\cC)$ can be obtained
by a slight extension of the results in \cite{WittDinst99}.}
.

 More explicitly, given a 2-cycle $\cC$ the `vector'
$m_I(\cC)$ projects on the components of $A^I_{[rs]}$ orthogonal
to $\cC$. This could be rephrased in more mathematical terms by
interpreting $A^I_{[rs]}$ and $\lambda^r_\cC$ as sections of
(non-trivial) holomorphic bundles \cite{WittDinst99}. In
particular, for $\cC\approx CP^1$,  decomposing $V\vert_\cC$ as
\be V\vert_\cC = \sum_{i=1}^{16} [\cO(k_i) \oplus \cO(-k_i)] \ee
and tensoring with the spin bundle $S_L = \cO(-1)$, one finds $dim
Ker(\bar\partial_{S_L\otimes V}) = \sum_i k_i$. The integers
$k_i$, with $k_i\ge 0$ without loss of generality, are further
constrained by the condition $C_2(T) = C_2(V)$ on the second Chern
class, that amounts to $G_3=d F_3 = 0$, since there are no D5-branes
in the ${\bf T}^6/{\bf Z}_3$ orientifold.

Our analysis differs from \cite{WittDinst99, SilvWit}, in that we
consider explicitly the coupling of the zero-modes of
$\lambda^r_\cC$ to the massless matter fields $A^I_{[rs]}$ in the
open string spectrum. Even in the presence of a non trivial
restriction to $\cC$ of the vacuum gauge bundle $V$ one can thus
have non-perturbative effects that require a field dependent
pre-factor $\epsilon^{rspq} A^I_{rs} A^J_{pq}$.

Rigid two-cycles $\cC$ with $\sum_i k_i = 4$ yield, after
integrating over $\lambda$'s and $\Theta$'s, superpotential term
of the form \be W_m = \sum_\cC m_I(\cC)m_J(\cC)\epsilon^{rspq}
A^I_{rs} A^J_{pq} \ee that generate a supersymmetric mass term for
all the $A$'s. Choosing a (canonical) basis of 2-cycles $\{ \cC_a
\}$ one can expand $\cC$ accordingly, \eg $\cC= \sum_a n^a \cC_a$,
and replace the sum over $\cC$ with a sum over $n^a$. The
dependence on $\cC$ hides the dependence on the K\"ahler moduli
$T$'s and $Z$'s that determine the sizes of the two-cycles.
Multiple covers are related to multi-instantons and may require
further investigation to be properly incorporated.

Considerations of $U(1)$ invariance suggest that each power of the
(pre)factor $\epsilon^{rspq} A^I_{rs} A^J_{pq}$ should be
accompanied by a compensating factor of $F(Z',T)\exp(-Z/3)$, where
$F(Z',T)$ only depends (holomorphically) on the K\"ahler moduli
(untwisted or twisted) which are neutral (do not shift) under the
anomalous $U(1)$. This means that a mass term and a quartic term
could only be generated by `fractional' instantons. A term of the
form $\det(A\otimes A)$ would instead require a compensating
$F(Z')\exp(-Z)$ which can be accounted for by 'standard' ED1
instanton wrapping cycles in integral homology. Although we cannot
produce a fully convincing argument, we expect these fractional
branes to be allowed at the orbifold point, where the $U(1)$
appears, and to support the correct number of $\lambda$ zero-modes
so as to produce the powers of $\epsilon^{rspq} A^I_{rs} A^J_{pq}$
upon integration. Indeed, the $Z_3$ trapped flux in the twisted collapsed
cycles generates the necessary fractional action via the
$\int B\wedge C_2$ WZ coupling of an ED3 brane wrapping a twisted four-cycle.

ED1's that wrap cycles that do not include the
cycle dual to the democratic $Z$ can also contribute
superpotential terms of the form \be W_0(Z') = \sum_{n_a} g(n_a)
\exp(-\sum_a n^a Z'_a) \ee which we expect to be T-dual to the
non-perturbative superpotential generated by wrapped ED3 and carefully
studied in the context of toroidal orbifolds with $\Omega_3$ and
$\Omega_7$ projections in \cite{lustetal}. As mentioned in the
introduction, the conclusion of \cite{lustetal} is that such a
superpotential combined with a flux superpotential and gaugino
condensation on D7 branes can completely stabilize the closed
string moduli as well as (some of) the open string moduli. Stable
uplift to dS, \ie a positive definite square mass in the AdS
ground-state, is only possible when complex structure deformations
are allowed. In particular this seems to exclude the ${\bf
T}^6/{\bf Z}_3$ case we focus on here. Yet inclusion of Scherk
Schwarz torsion and non-geometrical fluxes, that we will discuss
momentarily, and the non-perturbative superpotential discussed
above allows more possibilities.

{}{}{}{}{}{}{}{}{}{}{}{}{}{}{}{}{}{}{}{}{}{}{}{}{}{}{}{}{}{}

\section{Fluxes}
\label{fluxedz3}

We will now consider the possibility of turning on closed string
fluxes in the ${\bf Z}_3$ orbifold. Compatibly with the
orientation projection and barring $Z_2$ valued fluxes and open
string magnetic fluxes, the only available fluxes are the R-R
3-form flux along $Re \Omega$ or $Im\Omega$ (the real and
imaginary parts of the holomorphic 3-form) and the Scherk-Schwarz
torsion (metric fluxes). The flux superpotential is given by
\be
W_{flux} = \int (G_3^{R-R} - i \cT \circ J_C+{\cal R} \bullet (*S)) \wedge
\Omega^{CY}_3
\ee
where $*S$ is the 6-dimensional dual of the dilaton 0-form and the action of the
non-geometric flux is defined as
\be
\left({\cal R} \bullet A_p\right)_{j_1\cdots j_{p-3}}\equiv
R^{i_1i_2i_3}A_{i_1i_2i_3j_1\cdots j_{p-3}}\;\;.
\ee

{}{}{}{}{}{}{}{}{}{}{}{}{}{}{}{}{}{}{}{}{}{}{}{}{}{}{}{}{}{}

\subsection{Scherk-Schwarz Torsion on ${\bf Z}_3$ orbifold}

Let us denote the geometric torsion by $\cT$. In a real basis it
has components $\cT_{ij}{}^{k}$, with $i,j,k=1,...6$. In a complex
basis, for compatibility with the ${\bf Z}_3$ projection, it can
only have components $\cT_{IJ}{}^{\bar K}= - \cT_{JI}{}^{\bar K}$
and $\bar \cT_{\bar I\bar J}{}^{K}(\cT_{IJ}{}^{\bar K})^*$, with
$I,J, \bar K=1,2,3$.

The trace condition $\cT_{ij}{}^{i}=0$ is trivially satisfied by
the allowed components.

The cocycle condition \be \cT_{ij}{}^{l} \cT_{kl}{}^{m} + {\rm
cyclic \: in} \: (ijk) \: = 0 \ee imposes the following
constraints \be \cT_{IJ}{}^{\bar L} \bar \cT_{\bar K\bar L}{}^{M}
= 0 \ee \ie there is no further allowed cyclic permutation of the
lower complex indices.

Looking at $\cT_{IJ}{}^{\bar L}$ and $\bar\cT_{\bar K\bar
L}{}^{M}$ as 3 complex $3\times 3$ matrices \ie
$\cT_{IJ}{}^{\bar L} = (\cT_{I}){J}{}^{\bar L}$, $(\bar
\cT_{\bar I}){\bar J}{}^{K}$, the constraints read \be
(\cT_I)(\bar \cT_{\bar J}) = 0 \ee Moreover antisymmetry,
\ie $\cT_{IJ}{}^{\bar K}= - \cT_{JI}{}^{\bar K}$, implies
the $I^{th}$ row of matrix $\cT_I$ has all zero components and
the $J^{th}$ row of matrix $\cT_I$ has opposite components
w.r.t. the $I^{th}$ row of matrix $\cT_J$. Starting with the
diagonal constraints (no sum over $I$) \be (\cT_I)(\bar
\cT_{\bar I}) = 0 \ee one finds the following parametrizations
for the non vanishing rows of say $\cT_1$
\be
(\cT_1)_2{}^{\bar K} = (y_1, a_1 x_1, -x_1)\sp
(\cT_1)_3{}^{\bar K} = (\bar a_1 y_1, |a_1|^2 x_1, -\bar a_1
x_1)
\ee
where $x_1, y_1, a_1$ are three complex numbers.
Similarly
\be (\cT_2)_1{}^{\bar K} = (a_2 x_2,y_2,  -x_2) \sp
(\cT_2)_3{}^{\bar K} = (|a_2|^2 x_2, \bar a_2 y_2,  -\bar a_2
x_2)
\ee
and finally
\be
(\cT_3)_1{}^{\bar K} = (a_3 x_3,
-x_3,y_3)
\sp
(\cT_3)_2{}^{\bar K} = (|a_3|^2 x_3, \bar a_3 y_3,  -\bar a_3
x_3) \ee

Imposing antisymmetry one can relate $x_I, y_I, a_I$  with $I=2,3$
to one another
 and to $x_1, y_1, a_1$, that are not constrained any  further
 and can be used to parametrize the full solution. Dropping
 the index 1 for simplicity and setting $y = b x$ we find

\be (\cT_1)_2{}^{\bar K} = x (b, a , -1) \sp
 (\cT_1)_3{}^{\bar K} = x (\bar a b, |a|^2, -\bar a) \sp (\cT_2)_1{}^{\bar K} = - x (b, a , -1)
\ee
\be
(\cT_2)_3{}^{\bar K} = - x (|b|^2, a \bar b,  -\bar b) \sp
(\cT_3)_1{}^{\bar K} = - x (\bar a b, |a|^2, -\bar a) \sp
(\cT_3)_2{}^{\bar K} = x (|b|^2, a \bar b,  -\bar b)
\ee
that  satisfy the off-diagonal ($I\neq J$)
constraints, too.

The induced superpotential (in the $D9$-brane description) reads
\be W_{{\cal T}} = {Vol({\bf T}^6)\over 3} \epsilon^{IJK} \cT_{IJ}{}^{\bar K}
J_{K\bar K} \ee and depends on all 9 untwisted complex K\"ahler
parameters $J_{K\bar K}$,  in fact it is simply a linear
combination thereof (apart from the overall volume factor). In
principle one can also consider turning-on torsion in the
`twisted' sector that would induce a dependence of the flux
superpotential on the twisted K\"ahler moduli. Compatibility with
the non vanishing magnetic flux in the open string sector remains
to be investigated.

\section{Concluding remarks}
\label{conclusion}

We have derived the form of the non-perturbative and flux
superpotentials for Type I strings on the ${\bf Z}_3$-orbifold
after Chan-Paton symmetry breaking from $U(12) \times SO(8)$ to
$U(4)_{fp} \times U(4)^3$ (discrete Wilson lines) or $U(4)_{fp}
\times U(4)$ (continuous Wilson lines).

The determination of the precise numerical coefficients in front
of the non-perturbative terms would require a very detailed
analysis which is beyond the scope of the present investigation.
We don't expect the qualitative structure of the non-perturbative
terms to significantly change. In the case under consideration, an
ADS-like superpotential is generated by $ED5$, reproducing
`standard' gauge instantons. Additional `mass' terms are generated
by $ED1$, that represent `new' genuine stringy instantons. Closed
string fluxes generate additional terms.

Before attempting a full extremization of the complete
superpotential, possibly including perturbative terms that involve
matter charged under the `superconformal' $U(4)^3$ or $U(4)$, one
has to verify compatibility of the flux superpotential and the
`instanton' superpotential. Indeed, fluxes induce non trivial
warping of the geometry that may result in a `destabilization' or
`disappearance' of the cycles wrapped by $ED1$'s and of associate
non-perturbative terms. The ADS-like superpotential, due to
wrapped $ED5$ seems more robust, relying `only' on the compactness
of the internal manifold. It is tempting to conjecture that the
combined effect of $ED5$ and $ED1$ can stabilize the open string
`moduli' and the flux superpotential can then stabilize the closed
string moduli. We have only explicitly considered SS torsion in
the untwisted sector but it should not be impossible to consider
the effect of SS torsion in the twisted sectors.

We plan to address this and related issues in a forthcoming
investigation \cite{BKVZ}. In particular one should also analyze
the Fayet-Iliopoulos D-term for the anomalous $U(1)$ that should
roughly read $D = ReZ - 2 A^{\dagger} A$ and try to estimate the
corrections to the K\"ahler potential, that enters the expression
of the potential \cite{Dterms}. In addition to supersymmetric
extrema one could in fact hope to find non-supersymmetric dS
(meta)stable configurations.

In a series of papers \cite{lustetal},
superpotentials induced by fluxes and non-perturbative effects
 were studied in the case of orientifolds of
toroidal orbifolds (including the ${\bf T}^6/{\bf Z}_3$ case at
hand) with $\Omega$3/$\Omega$7–planes and D3– and D7–branes. The
main conclusions were that after resolution of the orbifold
geometries all (closed string) moduli can be stabilized in AdS but
only very few examples, the ones with (untwisted) complex
structure moduli, admit a stable uplift to DS. The inclusion of
open string moduli accounting for D3/D7– brane positions, Wilson
line moduli and matter fields was also analyzed in \cite{lustetal}
but no full fledged string models with all tadpole conditions
satisfied were produced.

In the present paper we have tried to partially fill in this gap
and to show that the combined effect of fluxes and $ED$-brane
instantons may generate interesting superpotential terms whose
combined effect may well stabilize closed as well as open string
moduli. In this respect it is tantalizing to observe that, at
fixed closed string moduli, $W_{ED5}(A)$ grows for small $A$'s
while $W_{ED1}(A)$ grows at large $A$. This is admittedly very
preliminary. In order to argue for complete moduli stabilization
one has to perform a more detailed analysis that should also
settle the issue of compatibility of the fluxes with tadpole
cancellation. We leave this for future work \cite{BKVZ}, where we
hope to address the possibility of meta-stabilization along the
lines of the ISS proposal \cite{metastab} that admits several
realizations in string theory \cite{metastabstr}. Another, largely
unexplored contribution to the scalar potential, we plan to
consider in some detail, is the D-terms from $U(1)$ R-R gauge
bosons that are present in type I models with
$\Omega_{5/9}$-planes when $h_{2,1}^-\neq 0$ and in models
$\Omega_{3/7}$-planes when $h_{2,1}^+\neq 0$.

%%%%%%%%%%%%%%%%%%%%%%%%%%%%%%%%%%%%%%%%%%%%%%%%%%%%%%%%%%%%%%%%%%%%%%%%%%%%%%%%%%%%%%%%%
\vskip 2cm
\begin{flushleft}
{\large \bf Acknowledgments}
\end{flushleft}
\vskip 1cm

\noindent It is a pleasure to thank P.~Anastasopoulos, R. Blumenhagen, E. Dudas, F.~Fucito,
S.~Kovacs, D. Lust, J.~F.~Morales and G.~Rossi for discussions. Most
importantly we would like to thank G. Villadoro and F. Zwirner for
collaboration in early stages and numerous enlightening
discussions.

This work was supported in part by the CNRS PICS no. 2530 and 3059, INTAS grant
03-516346, MIUR-COFIN 2003-023852, NATO PST.CLG.978785, the RTN grants MRTN-CT-2004-503369, EU MRTN-CT-2004-512194, MRTN-CT-2004-005104 and
by a European Union Excellence Grant, MEXT-CT-2003-509661.

 \newpage
\appendix

\end{document}